\documentclass[preprint2,twocolappendix]{aastex702}
\shorttitle{}
\shortauthors{}

\usepackage{scalerel,stackengine}
\usepackage{float}
\stackMath
\newcommand\Bhat[1]{%
\savestack{\tmpbox}{\stretchto{%
  \scaleto{%
    \scalerel*[\widthof{\ensuremath{#1}}]{\kern-.6pt\bigwedge\kern-.6pt}%
    {\rule[-\textheight/2]{1ex}{\textheight}}
  }{\textheight}%
}{0.5ex}}%
\stackon[1pt]{#1}{\tmpbox}%
}
\newcommand{\bs}[1]{\boldsymbol{#1}}

\usepackage{amsmath}
\usepackage{amsfonts}
\usepackage{amssymb}
\usepackage{natbib}
\usepackage{graphicx}
\usepackage{cancel}
\usepackage{xcolor}
\usepackage{ulem}
\usepackage{enumitem}
\usepackage{float}
\usepackage{mathrsfs} 
\usepackage{txfonts}
\usepackage[toc]{appendix}
\defcitealias{PP98}{1998PhRvE..57...21P}
\journalinfo{}

\begin{document}

\title{Decay of space plasma turbulence downstream of bow shocks: MMS magnetosheath unbiased campaign observations}

\correspondingauthor{Victor Montagud-Camps}

\author[0000-0002-7848-9200]{Victor Montagud-Camps}
\affiliation{Department of Electromagnetism and Electronics, University of Murcia, Spain}
\affiliation{Departament de Física, Universitat de les Illes Balears (UIB), 07122 Palma, Spain}
\affiliation{Institute of Applied Computing \& Community Code (IAC3), UIB, Spain}
\email[show]{victor.montagud@uib.cat}

\author[0000-0002-5608-0834]{Petr Hellinger}
\affiliation{Astronomical Institute of the Czech Republic, Czech Academy of Sciences}
\email{petr.hellinger@asu.cas.cz}

\author[0000-0002-4459-8783]{Sergio Toledo-Redondo}
\affiliation{Department of Electromagnetism and Electronics, University of Murcia, Spain}
\email{sergio.toledo@um.es}

\author[0000-0003-4380-4837]{Andrea Verdini}
\affiliation{INAF - Osservatorio Astrofísico di Arcetri, Firenze, Italy}
\affiliation{Dipartamento di Física e Astronomia, University of Florence, Italy}
\email{andrea.verdini@unifi.it}

\author[0000-0002-5981-7758]{Luca Sorriso-Valvo}
\affiliation{Istituto per la Scienza e Tecnologia dei Plasmi, CNR, 87036 Rende, Italy}
\affiliation{Department of Electromagnetics and Plasma Physics, School of Electrical Engineering and Computer Science, \\KTH Royal Institute of Technology, Teknikringen 31, SE-11428 Stockholm, Sweden}
\email{lucasorriso@gmail.com}

\author[0000-0003-4168-590X]{Francesco Pecora}
\affiliation{Department of Physics and Astronomy, University of Delaware, Newark, Delaware, USA}
\email{fpecora@udel.edu}

\author[0000-0002-5702-5802]{Julia E. Stawarz}
\affiliation{School of Engineering, Physics, and Mathematics, Northumbria University, United Kingdom}
\email{julia.stawarz@northumbria.ac.uk}

\author[0000-0001-8478-5797]{Alexandros Chasapis}
\affiliation{Laboratory for Atmospheric and Space Physics, Boulder, Colorado 80303, USA}
\email{alexandros.chasapis@lasp.colorado.edu}

\author[0000-0002-7969-7415]{Emanuele Papini}
\affiliation{INAF - Osservatorio Astrofísico di Arcetri, Firenze, Italy}
\affiliation{Instituto di Astrofísica e Planetologia Spaziali, Rome, Italy}
\email{emanuele.papini@inaf.it}

\author[0009-0005-2649-4795]{Inmaculada F. Albert}
\affiliation{Department of Electromagnetism and Electronics, University of Murcia, Spain}
\email{inma.albert@um.es}

\author[0009-0002-0988-8707]{Aida Castilla}
\affiliation{Department of Electromagnetism and Electronics, University of Murcia, Spain}
\email{aida.castillat@um.es}



\begin{abstract}
Turbulence properties in the solar wind change with increasing heliospheric 
distance. Observations from different missions have shown how spectral indices 
converge towards a $f^{-5/3}$ spectrum in the inertial range as the solar wind 
reaches one astronomical unit, and how the cascade rate decays as the plasma 
moves away from the Sun. We evaluate whether these properties display similar 
trends in a smaller astrophysical system, that of a planetary magnetosheath. 
Using data from the 2023 Magnetospheric Multiscale (MMS) unbiased 
magnetosheath campaign, we determine the evolution of the spectral indices and 
the cascade rate along 15 inbound magnetosheath crossings. WIND observations 
permitted to monitor the state of the solar wind during the campaign and 
allowed to study the impact of the upstream plasma conditions on the evolution 
of the cascade rate downstream the bow shock. Our results show an increase of 
the cascade rate between two to three orders of magnitude after the plasma 
crosses the bow shock into the magnetosheath. Once in the magnetosheath,  
turbulence develops with increasing distance to the bow shock in a way similar 
to solar wind observations, despite the smaller size of the system. 
Incompressible Hall-MHD scaling-law 
estimates of the cascade rate show a 
clear power-law decay of the cascade rate in the magnetosheath, particularly 
for the cases with upstream 
slow wind. This suggests that the way energy is 
injected into the system plays a role in the decay of turbulence as the plasma 
moves away from the bow shock.

\end{abstract}

\keywords{Interplanetary turbulence(830) --- Space plasmas(1544) --- Magnetosheath(2381) --- Solar-terrestrial interactions(1473)}


\section{Introduction} \label{sec:intro}
The terrestrial magnetosheath, a region formed past the bow shock due to the 
deceleration and compression of the solar wind at its encounter with Earth's 
magnetosphere, is an ideal environment to study plasma physics. The large magnetic 
field and plasma density 
in this region are features that distinguish it from the pristine solar wind in the 
near-Earth environment. Furthermore, its proximity to Earth makes it 
more accessible for in-situ observations than other regions of space. In order to 
monitor the plasma conditions in this environment, measurements have been taken  
using single and  multi-spacecraft missions, such as Cluster 
\citep{2001AnGeo..19.1197E}, THEMIS \citep{2008SSRv..141....5A}, and the Magnetospheric Multiscale (MMS) mission \citep{burch_magnetospheric_2016}. 

In particular, MMS measurements in the magnetosheath have lead to striking new 
discoveries in plasma physics, like the first observation of 
an electron velocity distribution being reshaped by a standard (ion) reconnection event 
\citep{2016Sci...352.2939B}, or the first measurements of 
electron-only reconnection jets \citep{Yordanova2016,2018Natur.557..202P}. MMS 
observations in the magnetosheath have also 
been especially relevant in the study of turbulence properties, as they have allowed us 
to 
analyze intermittency down to ion and electron scales \citep{2018JGRA..123.9941C}, to 
show the wave-vector anisotropy of velocity and magnetic fluctuations down to sub-ion 
scales 
\citep{2019FrP.....7..184R,2021JGRA..12628780H}, to determine the properties of fluctuations in an 
electron-only reconnection interval \citep{2019ApJ...877L..37S}, to determine the topology of magnetic and velocity fluctuations at kinetic scales in incompressible plasmas \citep{2023JGRA..12831064Z}, to observe the anisotropy of the cross-scale energy flux \citep{2023PhRvL.131v5201P} and to measure the pressure-strain 
contributions to the energy transfer \citep{2018ApJ...862...32C,2024PhRvL.132w5201M}.

These results have been possible thanks to the high resolution of the instruments onboard 
the MMS spacecraft, which have allowed us to explore a broad range of parameters and spatial 
scales. However, this parameter space is constrained by the limitations in the telemetry of 
data. The downlink of high-resolution data intervals is manually selected by a member of 
the MMS scientific community, the \textit{scientist in the loop} (SITL) based on 
the observation of events with low-cadence mode measurements. The SITL marks for downlink the data intervals that feature 
clear examples of known physical phenomena. As a consequence, the MMS high-cadence database can 
be considered to be biased towards certain plasma conditions, leading to a skewed 
understanding of the events that take place in the magnetosheath. 
This limitation of the MMS database motivated the creation of the first observational 
campaign dedicated to collecting an unbiased sample of high resolution data intervals of Earth's magnetosheath.

The 2023 MMS unbiased magnetosheath campaign provides a uniform sampling of Earth's 
magnetosheath at high cadence, with a $33\%$ coverage for 15 magnetosheath inbound 
crossings. The data intervals that constitute the campaign dataset were selected without 
any selection criteria besides that the spacecraft had to be located inside the 
magnetosheath. 
In addition to offering 
an unbiased selection of data, the campaign aims to provide 
observations to study the evolution of plasma properties across the magnetosheath with each 
inbound crossing. The data from this campaign has already been used to analyze the 
magnetosheath plasma 
properties in the unbiased dataset and their relation to intermittency \citep{2026JGRA..13134919P}, 
quantify  the occurrence of whistler waves in the magnetosheath \citep{2024JGRA..12932661S}, and 
determine the change of electron heat flux at different bow shock distances 
\citep{2025JGRA..13033272S}. In this work, we 
focus on analyzing the properties of plasma turbulence in 
this dataset, particularly the spectral indices and the energy cascade rate. Our aim is to 
present these turbulence properties in the unbiased magnetosheath dataset and 
determine whether they are affected by the distance to the bow shock.

Turbulence properties present a clear evolution with heliospheric distance in the inner heliosphere.
Statistical studies have shown a change of the power spectral density (PSD) of magnetic 
fluctuations above the ion Larmor radius, the so-called inertial range, from a power-law dependence in 
frequency of $f^{-3/2}$ to a steeper one $f^{-5/3}$ \citep{2020ApJS..246...53C,2023ApJ...943L...8S}. 
In fast winds and Alfvénic slow winds, low frequencies also display a $f^{-1}$ power law, whose 
breakpoint with the inertial range recedes to smaller frequencies as the heliospheric distance increases 
\citep{1982JGR....87.3617B,2013LRSP...10....2B}. 
Below the ion gyroradius, the spectra reaches the so called kinetic or sub-ion scales. There, the spectra present a steeper decay in frequency, with power-law 
exponents between -2 and -4, and a peak generally about 
-2.8 for observations made at one astronomical unit (AU) \citep{2006ApJ...645L..85S,2012ApJ...760..121A,2013ApJ...777...15S}. At electron scales, the power spectral density of magnetic fluctuations displays an even faster decay  with frequency \citep{2009PhRvL.103p5003A,2013ApJ...777...15S,2014ApJ...789L..28H,2017MNRAS.466..945M}.
In contrast with its evolution at inertial range scales, the spectra in the sub-ion range 
become less steep with heliospheric distance \citep{2020ApJS..246...55D,2022A&A...668A.189P,2025ApJS..278...40S}.
The measurement of spectral indices in the 
magnetosheath and their location with respect to the bow shock also suggest that turbulence follows a 
similar evolution in this environment: a steepening towards Kolmogorov's $f^{-5/3}$ at inertial range 
scales as the plasma moves away from the bow shock and the raise of the kinetic range power law, reaching $f^{-2.8}$ close to the 
magnetopause boundary 
\citep{huang_existence_2017,2020ApJ...898L..43L}. 
The 2023 unbiased magnetosheath campaign allows to verify, down to sub-ion scales, whether 
this statistical results hold true for and unbiased dataset and for individual 
magnetosheath crossings.

Spectral indices can be associated to a particular turbulence phenomenology, which in turn can serve to infer the rate at which nonlinear interactions transfer energy across scales  \citep{1991RSPSA.434....9K,1965PhFl....8.1385K}. This cross-scale energy transfer rate is also known as the energy cascade rate. There are ways to compute the energy cascade rate that do not rely on the assumption of a particular turbulence phenomenology. The Karman-Howarth Monin (KHM) equation 
\citep{1938RSPSA.164..192D,monin2013statistical} is a 
scale-dependent energy balance equation directly derived from the equations that model the fluid dynamics. In 
plasma turbulence, the KHM equation was initially obtained from the 
incompressible magnetohydrodynamics (MHD) equations by \citep{1998GeoRL..25..273P} (PP98) and it has since been 
extended to cover more plasma properties, including compressibility \citep{2013PhRvE..87a3019B,2021ApJ...917..101H}, expansion \citep{2013JGRA..118.1351H}, 
pressure anisotropy \citep{2022PhRvE.105e5111S}, pressure-strain interaction \citep{2022ApJ...929..142Y} and more \citep{2023PhR..1006....1M}. The use of the KHM equations has allowed the measurement of 

the cascade rate and its contributions in many  environments within the 
heliosphere, from the solar wind to planetary magnetosheaths \citep{2007PhRvL..99k5001S,2018PhRvL.120e5102H,2023ApJ...945....8A}. These measurements 
have revealed 
that not only the spectral 
indices change with heliospheric distance. Observations from Parker Solar Probe have shown 
that close to the Sun, the cascade rate is several orders of magnitude larger than at one AU \citep{2020ApJS..246...48B,2021ApJ...919...19A}. Helios observations at different locations along the heliospheric radial also indicate that the cascade rate decay with distance fits a power law whose exponent depends on the solar wind plasma properties \citep{Sorriso-Valvo2023}. We use the 2023 MMS unbiased 
magnetosheath campaign dataset to determine whether the evolution of the energy cascade rate and the spectral indices  
follow the same trend observed in the solar wind as 
the plasma in the magnetosheath moves  away from the bow shock. In order to achieve this goal, we compute the energy cascade rates in each high-cadence time interval of the unbiased campaign with the KHM equations derived from incompressible Hall-MHD by \citet{hellinger_von_2018} and 
\citet{ferrand_exact_2019}.

In section \ref{sec:KHM}, we have detailed 
the KHM equations for incompressible Hall-MHD. A detailed description of the 2023 MMS 
unbiased magnetosheath campaign (from now on, unbiased campaign) dataset can be found in section \ref{sec:data}. The results of 
our analysis are explained in section \ref{sec:results}, discussed in 
section \ref{sec:discussion} and summarized in section \ref{sec:conclusions}. 

\section{KHM equations for incompressible Hall-MHD} 
\label{sec:KHM}
Thanks to the high-resolution data collected by MMS during the unbiased campaign, we are 
able to analyze turbulence in the magnetosheath down to sub-ion scales, where kinetic 
effects such as the Hall current become relevant for the transfer of energy. In order 
to account for these contributions in the computation of the energy cascade rate, we use 
the KHM equation derived from the incompressible Hall-MHD equations:
\begin{eqnarray}
\partial_{t}\bs{u}+(\bs{u}\cdot\nabla)\bs{u}+\frac{1}{\rho_0}
\nabla\left(P+\frac{b^2}{2}\right)-\bs{b}
\cdot\nabla\bs{b}=\nu\Delta\bs{u}\\
\partial_{t}\bs{b}+(\bs{u}\cdot\nabla)\bs{b}-\bs{b}\cdot\nabla\bs{u}+d_i(\bs{b}\cdot\nabla)\bs{j}-d_i(\bs{j}\cdot\nabla)\bs{b}=\eta \Delta\bf{b}
\end{eqnarray}

where  $\boldsymbol{b}=(\boldsymbol{B}-\langle\boldsymbol{B}\rangle)/
\sqrt{\mu_0\rho_0}$ and $\bs{j}=\nabla\times\bs{b}$ are respectively the magnetic 
field fluctuation vector and the electric current in Alfvén speed units. The ion 
inertial length is denoted as $d_i$, the kinematic viscosity as $\nu$ and the 
magnetic resistivity as $\eta$.

Following the derivation by \citet{hellinger_von_2018} and 
\citet{ferrand_exact_2019}, the corresponding KHM equation for energy reads
\begin{eqnarray}
-\frac{1}{4}\frac{\partial \mathcal{S}}{\partial t}+
K-D=\varepsilon
\label{eq_KHM_hall}
\end{eqnarray}

Where $S(\bs{\ell})=\left\langle 
|\delta \boldsymbol{u}|^2+|\delta \boldsymbol{b}|^2 \right\rangle$ is the 
second-order 
structure function, $\delta\boldsymbol{a}(\bs{\ell},\boldsymbol{x})\equiv 
a(\bs{x}+\bs{\ell})-a(\bs{x})$ is the increment across the spatial lag $\bs{\ell}$ for a generic 
quantity $\bs{a}$, $\boldsymbol{u}$ and $\boldsymbol{b}$ are the velocity fluctuations and magnetic 
field fluctuations in Alfvén speed units, and $\langle\dots\rangle$ is the spatial average over the domain.

The first term on the left-hand side of equation \ref{eq_KHM_hall} 
corresponds to the temporal 
variation of the energy of the fluctuation at a scale $\ell$. The contribution to the cascade 
associated to nonlinear interactions is given by the second term $K(\bs{\ell})$. It is the 
sum of the contribution derived from the nonlinear terms in incompressible MHD,  
$K_{MHD}(\bs{\ell})=-\frac{1}{4}\nabla\cdot\bs{Y}$, and the contributions 
associated to the Hall term, 
$K_{H}(\bs{\ell})=-\frac{1}{4}\nabla\cdot\bs{H}$. The Yaglom and Hall vector 
fields, $\bs{Y}(\bs{\ell})$ and $\bs{H}(\bs{\ell})$, are defined as

\begin{eqnarray}
 \bs{Y}(\bs{\ell})= \left \langle \delta\bs{u}
 (|\delta\boldsymbol{u}|^2+|\delta\boldsymbol{b}|^2) 
-2\delta\boldsymbol{b}(\delta\boldsymbol{u}\cdot\delta\boldsymbol{b}) \right 
\rangle 
\\
 \bs{H}(\bs{\ell})= \frac{1}{2}d_i\nabla\cdot\left\langle
 2\delta\bs{b}(\delta\bs{b}\cdot\delta\bs{j})-\delta\bs{j}(\delta\bs{b}\cdot\delta\bs{b}) 
 \right \rangle
\end{eqnarray}

The dissipative contributions to the energy transfer are split into 
a scale-dependent and a scale-independent component. $D(\bs{\ell})$ is the scale-dependent term, defined as 
\begin{eqnarray}
 D(\bs{\ell})= -\frac{1}{2}\nu\Delta\left\langle |\delta\bs{u}|^2
 \right\rangle
-\frac{1}{2}\eta\Delta\left\langle|\delta\bs{b}|^2\right\rangle
\end{eqnarray}

and $\varepsilon=\nu\langle\boldsymbol{\nabla u}:\boldsymbol{\nabla u}\rangle+\eta\langle\boldsymbol{\nabla b}:\boldsymbol{\nabla b}\rangle$ is the scale-independent contribution 
to dissipation and corresponds to the energy cascade rate. Here, the operator $:$ denotes the double contraction of two second-order tensors.

In order to evaluate each term of the KHM equation using 
observations, we assume i) that the temporal variation and scale-dependent 
dissipative KHM terms have a negligible contribution to the energy 
cascade in the range of scales that we are measuring \citep{hellinger_von_2018}; 
ii) that Taylor's hypothesis \citep{1938RSPSA.164..476T} is fulfilled in that range of scales and therefore $\ell=U_{sw}\tau$, where $U_{sw}$ is the bulk wind speed 
and $\tau$ is the temporal lag; and iii) that turbulence is 
isotropic. From these assumptions, it follows the scaling law \citep{2009EL.....8825001C}

\begin{eqnarray}
\varepsilon
\approx
-\frac{3}{4}\frac{Y_R(\tau U_{sw})+H_R(\tau U_{sw})}{\tau U_{sw}}
\label{eq_exact_law}
\end{eqnarray}

where the R subindex denotes the component of the vector fields in the flow 
direction. Since we assume turbulence isotropy, we take $Y_R+H_R\approx Y_x+H_x$, where the subindex x refers to the x direction in Geocentric Solar Earth (GSE) coordinate system.

In a well-developed turbulent flow, the KHM terms in the right-hand side of equation 
\ref{eq_exact_law} should be constant and equal to the cascade rate, $\varepsilon$, 
across the inertial range and Hall-related scales. Conversely, when turbulence is not 
fully developed or when some of the aforementioned assumptions are not fulfilled, strong variations on the right-hand side of equation 
\ref{eq_exact_law} take place across the scales. We only assign a  
cascade rate value to those unbiased campaign intervals where the scale-dependent terms in equation \ref{eq_exact_law} remain  approximately constant. We select such intervals 
by visual inspection. For the intervals where 
the absolute value of the KHM terms \textit{converge} to a 
relatively constant value, the cascade 
rate is computed as the average of the right-hand side terms in equation \ref{eq_exact_law}
between the ion inertial length scale and the correlation length 
$\ell_{corr}=\tau_{corr} U_{sw}$. The correlation time, $\tau_{corr}$, is computed 
using the normalized autocorrelation of magnetic field fluctuations 
$R(\tau)=\left\langle\bs{b}(t+\tau)\cdot\bs{b}(t)\right\rangle/\left\langle\bs{b}(t)^2\right\rangle$. Assuming that 
$R(\tau)$ decays exponentially, the correlation time can be defined as 
$R(\tau_{corr})=e^{-1}$ \citep{1998JGR...103.6495M}. 
In order to estimate this e-folding correlation time, the correlation function 
$R(\tau)$ is computed for each time interval and fit to an exponential function. In 
this work, the correlation time and the cascade rate obtained from equation 
\ref{eq_exact_law} are computed in each MMS spacecraft and then averaged over the four 
spacecraft.

\section{Data selection} 
\label{sec:data}
\begin{figure}
\includegraphics[width=.45\textwidth]{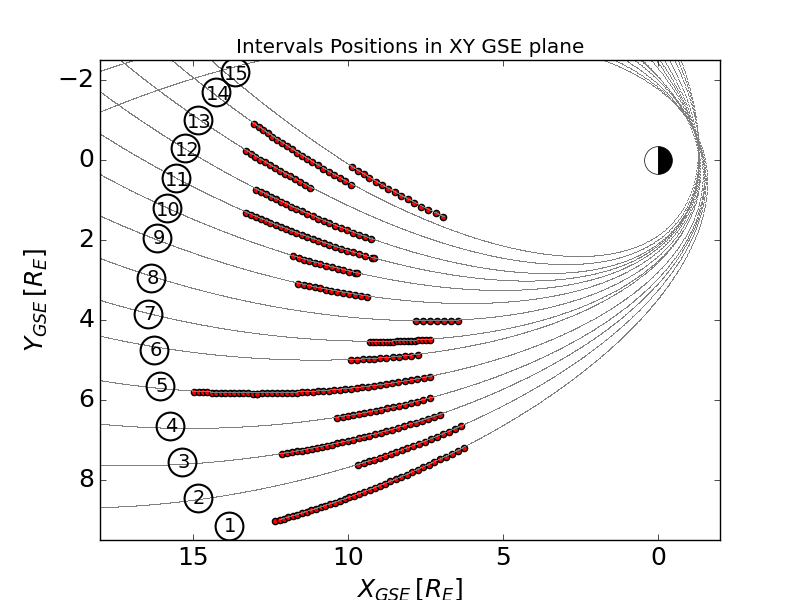}
\caption{Spatial distribution of each 3-minute burst intervals during 
the 2023 unbiased magnetosheath campaign. The red circles correspond to the position of MMS1 
projected onto the XY Geocentric Solar Ecliptic plane at the middle of each 3-minute burst 
interval. Grey lines represent the orbit of MMS1 during the campaign. Magnetosheath crossings are numbered from 1 to 15, starting from the first crossing of the campaign. The same numbering is used in table \ref{table_1}.
}
\label{fig_orbit}
\end{figure}

The Magnetospheric Multiscale mission (MMS) consists of four identical spacecraft that collect in-situ data in the near-Earth environment. Between 2 February 2023 and 24 March 2023, 
MMS gathered 309 data intervals at high cadence (burst mode) from the magnetosheath in the context of the 2023 unbiased magnetosheath campaign. For each of these time intervals, figure \ref{fig_orbit} shows the 
average position of MMS1 projected onto the Geocentric Solar Ecliptic XY plane. 
The time intervals were collected during 15 inbound crossings. Each interval  
has a duration of 3 minutes and they are separated from each other by a 6 minute gap in 
burst mode data. For each inbound orbit, the first 3-minute burst interval is taken right 
after the bow shock, and 
the last one is collected right before the spacecraft reach the magnetopause. The bow shock and magnetopause locations  
were determined by visual inspection by a scientists-in-the-loop (SITLs) in charge of the campaign. Besides these conditions, no other criteria were applied for the selection of the intervals.

For our analysis, we use the plasma moments from the second-level data products of the 
Fast Plasma Investigation (FPI) instrument \citep{pollock_fast_2016}. In 
burst mode, FPI provides a measurements of the plasma moments for 
ions each 150 ms and for electrons each 30 ms. 
We also use the high-cadence magnetic field data from the Fluxgate Magnetometer 
(FGM) instrument \citep{2016SSRv..199..189R}, which has a resolution of 
$128$ Hz in burst mode. Since the four MMS spacecraft maintained a  
tetrahedron flight formation during the unbiased campaign measurements, we compute the 
current density with the curlometer technique \citep{2002JGRA..107.1384D} 
using the magnetic field data from all four spacecraft.

In addition to the MMS data, we also use measurements from the WIND spacecraft at 1 AU in order 
to evaluate the conditions upstream the bow shock during the unbiased campaign. We take 
time intervals with WIND starting one hour before each magnetosheath inbound crossing. This 1-hour delay is an estimation of the time it takes the solar wind to propagate from WIND's location to the bow shock. Each WIND interval has a duration of 8 hours, in order to measure the turbulence correlation time in the solar  wind at 1 AU, which typically display values between one to three hours \citep{1982JGR....8710347M,2013JGRA..118.3995W,2013LRSP...10....2B}, that is, longer than some of the magnetosheath traversals of the unbiased campaign.
From these WIND intervals, we collect the velocity of the plasma from the 3DP/PESA instrument \citep{1995SSRv...71..125L}, which has a maximum 
resolution of 3 seconds. The magnetic field data is obtained from the Magnetic Field Investigation (MFI) instrument \citep{1995SSRv...71..207L} and is resampled to 3DP/PESA resolution using linear interpolation. 

\section{Results} 
\label{sec:results}

\subsection{Power spectral density} 
\label{subsec:psd}

We have computed three characteristic scales and their 
corresponding frequencies for each 3-minute burst interval: the correlation length, the proton gyroradius, $\Omega_i$, and the 
ion inertial length, $d_i$. The 
trace of the power spectral tensor
of magnetic field fluctuations is obtained for each 3-minute 
burst interval. In order to compute the spectral indices in the inertial range, we fit the spectra to a power law 
of frequency, $f^{\alpha}$, between the frequency associated to the correlation length and the ion gyrofrequency. 
For the sub-ion range, we fit the spectrum between the ion gyrofrequency 
and 10 Hz, which provides a wide-enough range to compute the kinetic spectral index, generally about a decade-long frequency range.

The boundaries of the intervals where the inertial and kinetic spectral indices are computed are taken differently when the burst intervals present a bump in the spectrum. 
Such local enhancements of the spectrum are associated to the presence of 
certain waves or structures in the plasma, like mirror modes or Alfvén 
vortices \citep{2006PhRvL..96g5002S,2016ApJ...824...47L,2018JGRA..123.5300R}. 
After visual inspection of each spectra, we 
detected 24 time intervals of the unbiased campaign with these features. In most cases, the spectral 
enhancements appear near the transition region to kinetic scales, in a 
narrow range of frequencies. In this context, we set the frequency intervals to compute the inertial and 
kinetic spectral indices so that the bump is not included 
in them. Among the intervals with spectral enhancements, there are 15  
where the bump section dominate the frequencies smaller than the proton gyrofrequency, making it impossible to measure the spectral index at inertial 
range scales. Those 15 bump-dominated intervals are not considered in the 
statistics of the inertial range spectral indices.

\begin{figure}
\includegraphics[width=0.45\textwidth]{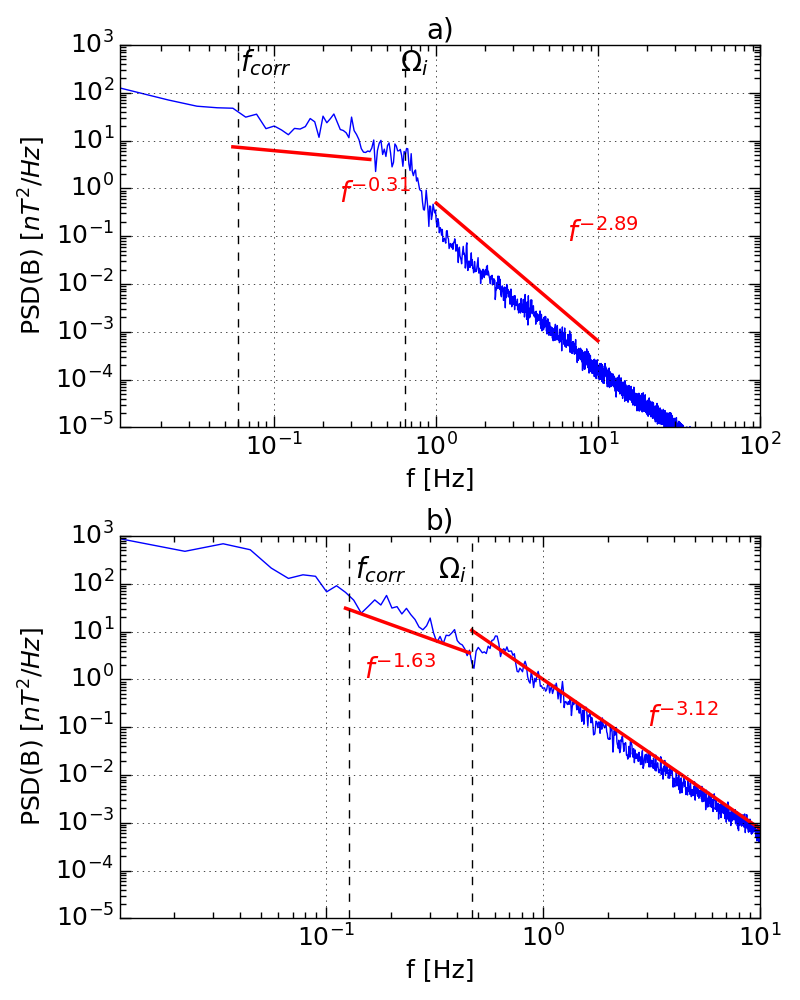}
\caption{ Examples of spectra of the magnetic field fluctuations for two 3-minute burst intervals of the 2023 MMS unbiased magnetosheath campaign. Red lines show the frequency power laws $f^{\alpha}$ that fit the spectra in the inertial and sub-ion frequency ranges. Panel a): example of flat inertial range measured in the time interval [2023-02-17 00:55:43, 2023-02-17 00:58:43] UTC. Panel b): example of Kolmogorov-like inertial range measured in the time interval [2023-02-16 22:31:53, 2023-02-16 22:34:53] UTC.
}
\label{fig_example_spectra}
\end{figure}

\begin{figure}
\includegraphics[width=0.45\textwidth]{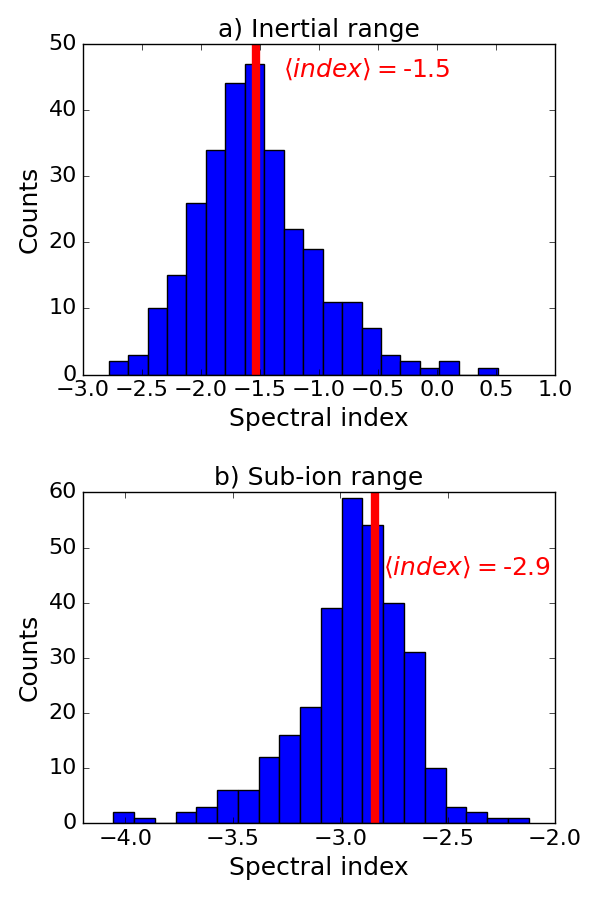}
\caption{Histograms of spectral indices computed for all burst intervals of the unbiased campaign. Panel a) corresponds to the spectral indices in the inertial range.  This histogram does not include the 15 intervals where the spectral index was not possible to compute due to the presence of a bump-dominated spectrum at inertial range scales. Panel b) is the histogram of spectral indices within the sub-ion scales. Red vertical lines mark the mean value in each histogram.
}
\label{fig_hist_spectra}
\end{figure}

Figure \ref{fig_example_spectra} presents 
the power spectral density (PSD) of two burst intervals 
and the fitted power laws for each frequency range. 
Panel \ref{fig_example_spectra}a) shows a power law with an index 
$\alpha=-0.31$ in the inertial range. The spectral index has been computed between 
the frequency associated to the correlation length, $f_{corr}$, and $f=0.3$ Hz, in order to avoid taking 
into account the effect of the spectral enhancement near $\Omega_i$. Panel \ref{fig_example_spectra}b), on the 
other hand, presents a power law with an index $\alpha=-1.63$ in the inertial range, that is, close to 
the $-5/3$ spectral index predicted by Kolmogorov for well-developed turbulence in hydrodynamics. The 
spectral indices at sub-ion scales are similar in both cases, with $\alpha=-2.89$ in 
panel \ref{fig_example_spectra}a) and 
$\alpha=-3.12$ in panel \ref{fig_example_spectra}b). Figure \ref{fig_hist_spectra} compiles into two histograms the result of computing the spectral indices 
for each 3-minute burst interval of the unbiased campaign. Panel 
\ref{fig_hist_spectra}a) shows that the inertial range spectral indices span approximately from $-2.8$ to $0$, with mean value at $-1.5$ and a distribution skewed towards the positive values, with 
skewness $0.9$. The histogram in panel \ref{fig_hist_spectra}b) shows that spectral indices at
are 
sub-ion scales take values between $-4$ and $-2.5$, and are more symmetrically distributed around its 
mean value at $-2.9$, with skewness of 
$-0.3$. These results are consistent with measurements of the sub-ion range spectral index in the solar wind at 1 AU and in previous observations in the magnetosheath 
\citep{2013PhRvL.111n9001A,2014ApJ...789L..28H,huang_existence_2017}.

\begin{figure}
\includegraphics[width=0.5\textwidth]{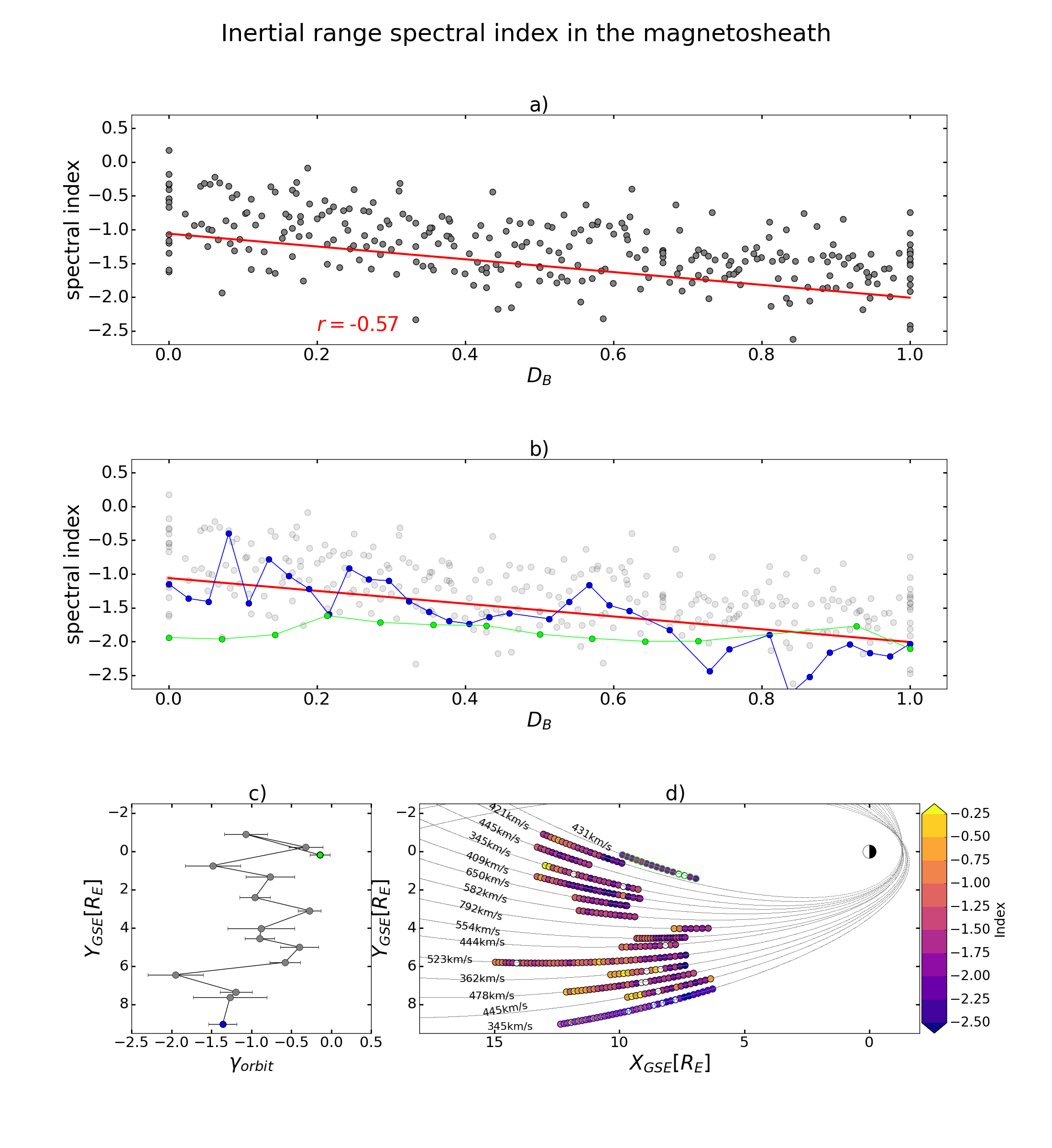}
\caption{Evolution of spectral indices in the inertial range with respect to the relative distance to the bow 
shock. The horizontal axis indicates the relative distance to the bow shock, $D_{B}$, 
where 0 corresponds to the position of the bow shock (BSH) crossing, and 1 marks the 
magnetopause (MP) crossing. Panel a): Inertial range spectral indices, $\alpha$, versus 
relative distance $D_{B}$.  Panel b): the same as a), but with the data points from 
the closest magnetosheath traversal to the flanks highlighted in blue (corresponding to crossing $1$ in figure \ref{fig_orbit}) and to the 
 closest to the subsolar region 
in green (crossing $15$). Red lines in panel a) and b) correspond to the linear fit $\alpha=\gamma D_{B}+n$ computed for all burst intervals, and the correlation 
coefficient is indicated as $r$ in panel a). Panel c): slope of the linear fit, $\gamma$, for each magnetosheath traversal versus 
the location in $Y_{GSE}$ of MMS1 during the first (furthest from Earth) burst interval of each inbound crossing. The blue and green points correspond to 
crossings 1 and 15, as indicated in figure \ref{fig_orbit}. Panel d): location of MMS1 in the XY GSE 
plane at the middle of each burst interval. The color scale indicates the 
inertial range spectral index of the burst interval. The first 
and last orbits are highlighted in blue and green. White circles 
correspond to intervals where it was not possible to compute the spectral 
index due to a bump-dominated region at inertial range scales. Velocities in the left of each magnetosheath crossing indicate the averaged upstream solar wind speed during each magnetosheath traversal. 
}
\label{fig_evol_inertial}
\end{figure}

In order to evaluate whether the distance to the bow shock has an 
effect on the slopes of the spectra, we estimate the relative 
position of each 3-minute interval with respect to the bow shock 
and the magnetopause, named $D_B$ from now on. Based on the selection criteria of burst intervals for the unbiased campaign, we  
consider the first timestamp of 
the first 3-minute interval in each inbound orbit  
as the one marking the instant of the bow shock crossing. 
Similarly, we take the last timestamp of the last burst interval 
of each magnetosheath traversal as the instant that the magnetopause 
was reached by MMS. Figure \ref{fig_evol_inertial} shows how the inertial range spectral 
indices are distributed in the magnetosheath and evolve with distance in each 
magnetosheath traversal. 
Panel \ref{fig_evol_inertial}a) presents the spectral indices in 
the inertial range and the relative position of MMS1 for all the burst intervals of the 
unbiased campaign. The red line corresponds to the linear fit of the spectral indices 
with $D_{B}$, which has a correlation coefficient of 
$r=-0.57$. This panel indicates 
that the inertial range spectral indices are flatter close to the bow shock. 
Approximately at the bow shock crossing, $D_B=0$, the average 
spectral index is $-1$. 
The spectra become steeper as the plasma is closer to the magnetopause. At $D_B=1$, 
which corresponds to the magnetopause crossing, the average spectral index 
becomes $-1.8$.

Panel \ref{fig_evol_inertial}b) highlights the evolution of the spectral 
indices for the first and the last orbit of the campaign, in blue and green, 
respectively. The first orbit of the campaign, which is the one closest to the 
magnetosheath flanks, shows a clear decrease of the spectral index with distance,
although with some occasional fluctuations along the magnetosheath 
traversal. Conversely, 
the spectral indices measured for the last orbit of the unbiased campaign, taken close to the subsolar region, remain relatively constant 
across the magnetosheath, with spectral indices generally about 
$-2$. 
Panel \ref{fig_evol_inertial}c) shows this dependence with the distance to the 
subsolar region for each magnetosheath crossing of the campaign. We computed the slope of the 
linear fit $\alpha_{inertial}=\gamma D_B+n$ for each magnetosheath traversal, 
where $\alpha_{inertial}$ is the inertial range spectral index. For each magnetosheath 
crossing, panel \ref{fig_evol_inertial}c) displays  
the $\gamma$ and the position of 
MMS1 in $Y_{GSE}$ at the bow shock crossing in Earth radii, $R_{E}$. The blue and green points represent the 
same orbits as in panel \ref{fig_evol_inertial}b). As the orbits approach the subsolar 
region, near $Y_{GSE}=0\,R_{E}$, panel \ref{fig_evol_inertial}c) shows that the spectra at the 
inertial range scales flatten less with increasing $D_B$. 
Finally, panel \ref{fig_evol_inertial}d) shows the position of MMS1 in the XY GSE plane 
at middle of each burst interval. At the left side of each MMS traversal, 
we indicate the average solar wind speed upstream the bow shock measured by WIND 
spacecraft. Each point in panel \ref{fig_evol_inertial}d) is 
colored according to its spectral index value in the inertial range. This panel shows  
the burst
intervals in each magnetosheath traversal and provides a 
qualitative view on the influence of the upstream conditions 
on the evolution of the inertial range spectral index across the magnetosheath. The 
traversals with highest solar wind speed upstream of the bow shock are also the 
shortest ones, which is indicative of a more compressed magnetosheath. No qualitative 
difference is observed in the evolution of the spectral indices of the shortest and the 
largest magnetosheath traversals.

\begin{figure}
\includegraphics[width=0.5\textwidth]{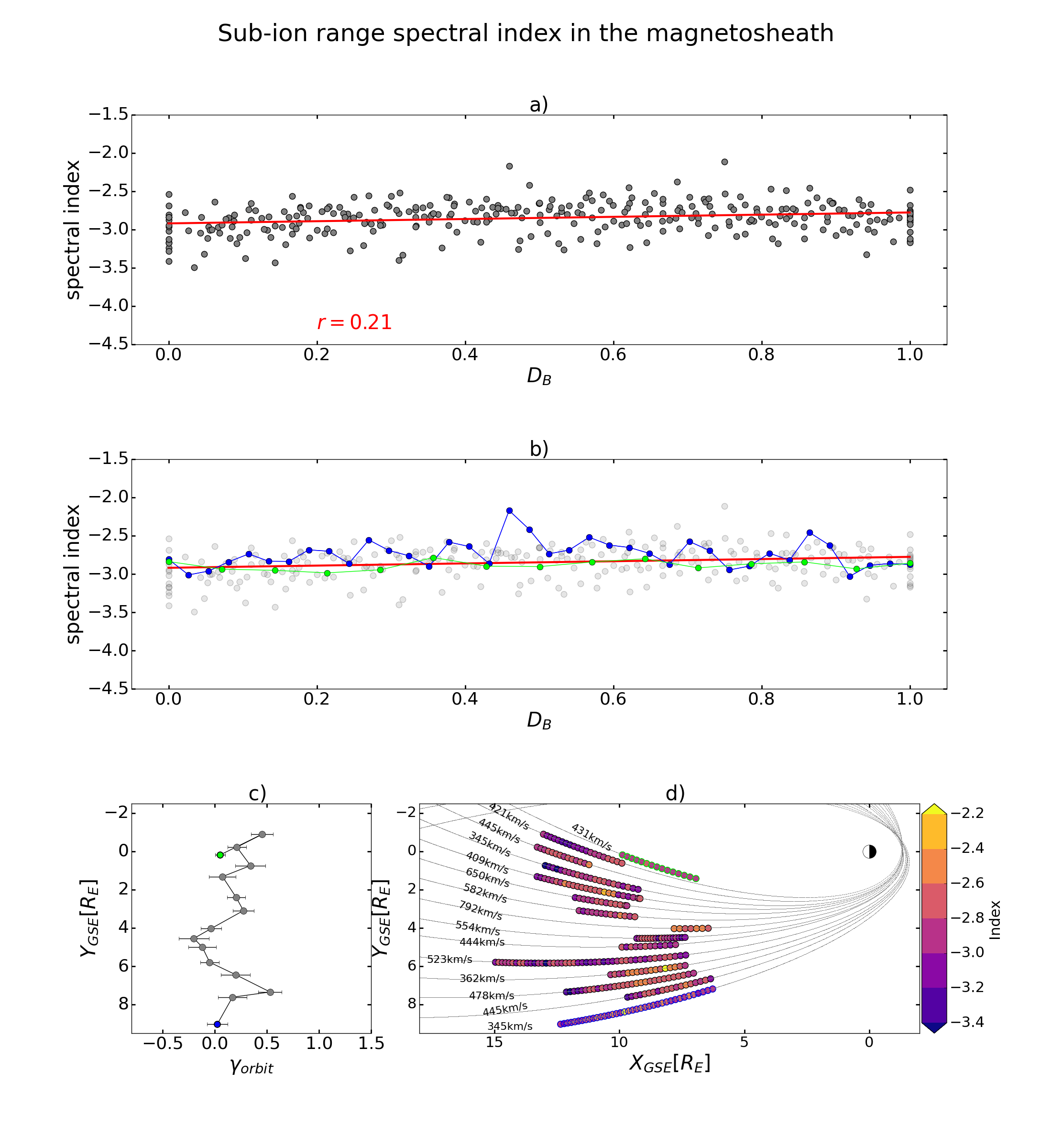}
\caption{Evolution of spectral indices in the sub-ion range with respect to the relative  distance to the bow shock. Panels depict the same information as in figure  \ref{fig_evol_inertial}.
}
\label{fig_evol_kinetic}
\end{figure}

Figure \ref{fig_evol_kinetic} provides the same information as figure \ref{fig_evol_inertial}, but  
for the sub-ion range spectral indices. Panel \ref{fig_evol_kinetic}a) shows that  
spectral indices measured in the sub-ion range for the whole campaign follow the 
opposite trend than the inertial range spectral indices. At sub-ion scales, the 
spectral indices increase, although slightly, as the spacecraft approaches the magnetopause. 
The spectral indices also fit a linear function of the form $\alpha_{kinetic}=\gamma D_B+n$, which is depicted as a red line in panels \ref{fig_evol_kinetic}a) and 
\ref{fig_evol_kinetic}b). The smaller variation of the sub-ion spectral indices with $D_B$ leads to a correlation coefficient of 
$r=0.21$, smaller than 
for the inertial range scales. 
The points highlighted in panel \ref{fig_evol_kinetic}b) indicate again that 
for the nearest crossing to the subsolar region (green points), the spectral indices are 
closer to a value characteristic of a well developed turbulence and almost unaltered 
across the magnetosheath, while a clearer evolution can be observed for the 
magnetosheath traversal closer to the flanks (blue points). Panels  
\ref{fig_evol_kinetic}c) and \ref{fig_evol_kinetic}d) show the dependence with  
the distance to the subsolar region for each crossings. In panel \ref{fig_evol_kinetic}c), it 
can be observed that the slope of the linear fit is positive near the flanks and 
approaches $0$ as the location of the bow shock crossing also approaches $Y_{GSE}=0\,R_E$. The plasma speed 
also seems to play a role in the evolution of the sub-ion 
range spectral indices with distance.
The upstream bulk plasma speed appears to the left of each magnetosheath crossing in panel \ref{fig_evol_kinetic}d). The 
magnetosheath crossings with the largest upstream speeds, tend to have kinetic spectral indices that remain constant or decay with $D_{B}$. This suggests an influence of the solar wind on the sub-ion scale 
spectral indices in the magnetosheath, but a larger sample of upstream conditions is necessary to verify 
it.

\subsection{Energy cascade rate} 
\label{sec:cascade}

The nonlinear contributions of the incompressible Hall-MHD KHM, $K_{MHD}$ and 
$K_{Hall}$, and their sum, $K_{NL}$, are computed for all burst intervals of the 2023 unbiased magnetosheath campaign using the right-hand side of equation \ref{eq_exact_law}. Since we assume that in the range of scales measured in each burst interval the dissipative and non-stationary contributions of the KHM equation are negligible, the sum of the nonlinear terms is the only contribution 
to the cross-scale transfer of energy and should present a nearly constant value.

\begin{figure}
\includegraphics[width=0.45\textwidth]{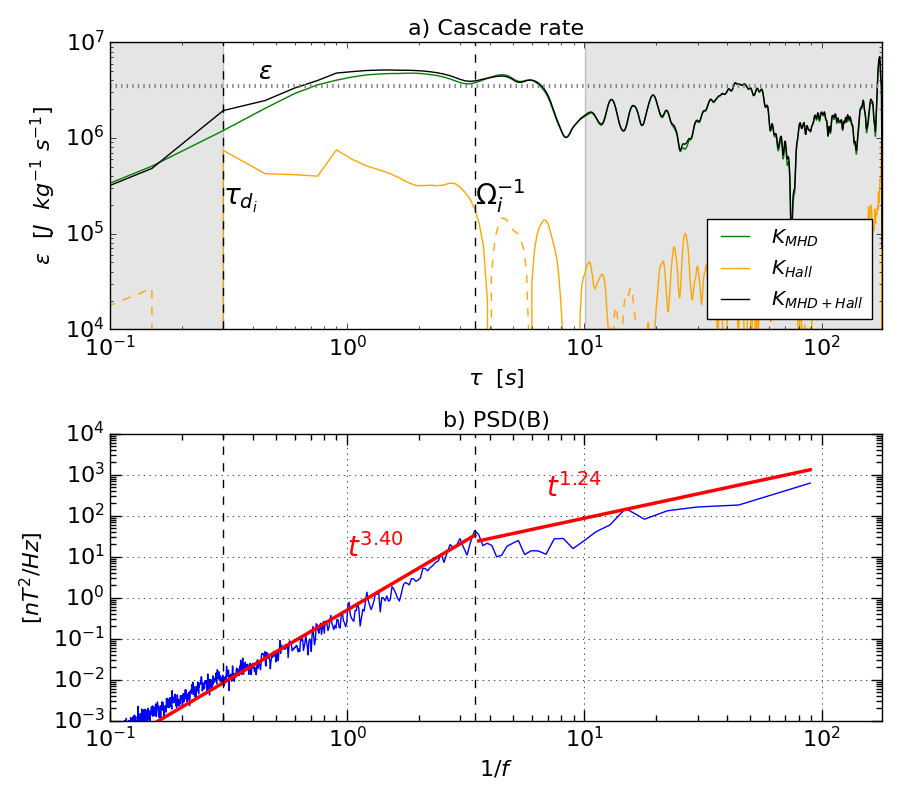}
\caption{Example of measurement of the cascade rate in a burst interval of the unbiased 
campaign, [2023-02-16 23:07:43, 2023-02-16 23:10:43] UTC. Panel a): contributions to the energy cascade rate: nonlinear MHD transfer 
$K_{MHD}$ (green); hall current contribution, 
$K_{Hall}$ (yellow) and sum of both (black). Shaded regions correspond to scales not used 
to compute the energy cascade rate, $\varepsilon$ (dotted horizontal line). Panel b): power spectral density of the  magnetic field fluctuations.
}
\label{fig_KHM_example}
\end{figure}

Panel a) of figure \ref{fig_KHM_example} shows an example of the nonlinear KHM terms computed for the burst interval [2023-02-16 23:07:43, 2023-02-16 23:10:43] UTC. The Hall contribution is 
depicted as a yellow line, the MHD contribution as a green one and the sum of both as a black line. We compute the energy cascade rate 
$\varepsilon$ as the unsigned average of the total nonlinear contribution in the non-shaded region, and we represent it in panel \ref{fig_KHM_example}a) with a horizontal dotted line. The non-shaded 
region is delimited by the correlation time of the burst interval and the time associated to the ion-inertial length. Hence, the region of scales where the energy cascade rate is estimated includes the inertial range and part of the sub-ion range.  Panel \ref{fig_KHM_example}b) shows that the spectrum changes within the 
range of scales where we compute the energy cascade rate. In this particular interval, the spectrum follows a $t^{1.24}$ power law in the inertial range, while it becomes $t^{3.4}$ for time scales  smaller than the inverse of the proton gyrofrequency $\Omega_{i}$.

We apply the same procedure to measure the energy cascade rate in all burst intervals 
of the unbiased campaign. Only the $\varepsilon$ computed in intervals that present 
small variations of $K_{NL}$ across scales are kept. 
By visual inspection, we find that 
152 out of the 309 initial burst intervals 
 satisfy this criterion. Since it is necessary to 
fulfill Taylor's hypothesis in order to compute the energy cascade rate using single-spacecraft measurements, we test the validity of the hypothesis over the previous selection of $152$ magnetosheath burst intervals. In 
order to do so, we keep the intervals that fulfill $U> v_A$, where 
$U$ and $v_A$ are 
respectively the average bulk plasma speed and Alfvén speed of each 
burst interval. As a result, we obtain $114$ burst intervals where 
the cascade rate can be measured, out of the initial initial $309$ 
intervals of the unbiased campaign. From now on, 
we name these $114$ intervals as \textit{convergent} and the rest of 
the burst intervals as \textit{divergent}.

In addition to the condition $U> v_A$, we also evaluate the validity of 
Taylor's hypothesis by testing whether multi-spacecraft measurements of statistical 
quantities taken at the separation scale between MMS spacecraft give the same 
result as single spacecraft measurements \citep{2017ApJ...842..122C,2018JGRA..123.9941C,2019ApJ...877L..37S,2022PhPl...29a2302S}. The quantity we use to make this comparison 
between single and multi-spacecraft measurements is the second order structure function 
of the magnetic field fluctuations $S_b(\ell)=\langle|\delta\bs{b}(\ell)|^2\rangle$ at 
the mean separation between MMS spacecraft during each burst interval, 
$\ell_D$. This distance can be directly measured for all six MMS 
spacecraft 
pairs, denoted as $ij$, and can be used to compute the second 
order structure function $S_{b,ij}(\ell_D)=\langle|\vec{b}_i(\vec{r}_{i})-\vec{b}_j(\vec{r}_{j})|^2\rangle_t$, 
where $\vec{r}_{i}$ and $\vec{r}_{j}$ are the positions of the spacecraft pairs, $\vec{b}_i$ and $\vec{b}_j$ are the magnetic field fluctuations in Alfvén speed units measured by each spacecraft, $\ell_D=|\vec{r}_{i}-\vec{r}_{j}|$ is the spacecraft separation and $\langle\cdots\rangle_t$ is the time average over the 3-minute burst interval. On the other hand, assuming Taylor's hypothesis is valid for each MMS spacecraft, the single spacecraft second order structure function is defined as $S_{b,i}(\ell_D)=\langle|\vec{b}_i(U\cdot(t-\tau_D)))-\vec{b}_i(U\cdot t)|^2\rangle_t$. The ratio
 $r_T=\langle S_{b,i}\rangle_i/\langle S_{b,ij}\rangle_{ij}$ is measured for 
each burst interval, where $\langle\cdots\rangle_i$ and $\langle\cdots\rangle_{ij}$ are the average over the four MMS spacecraft and the average over all six spacecraft pairs. The ratio  
$r_T$ takes values between $0.12$ and $2.3$ and its average for all burst intervals is  
$0.7\pm 0.4$. Following \citet{2022PhPl...29a2302S}, we consider that Taylor's hypothesis is well satisfied if the ratio is a factor 2 from unity, $r_T\in[0.5,2]$. All $114$ convergent intervals fulfill this condition, while $57$ of the divergent intervals present ratios outside this range.

\begin{figure}
\includegraphics[width=0.55\textwidth]{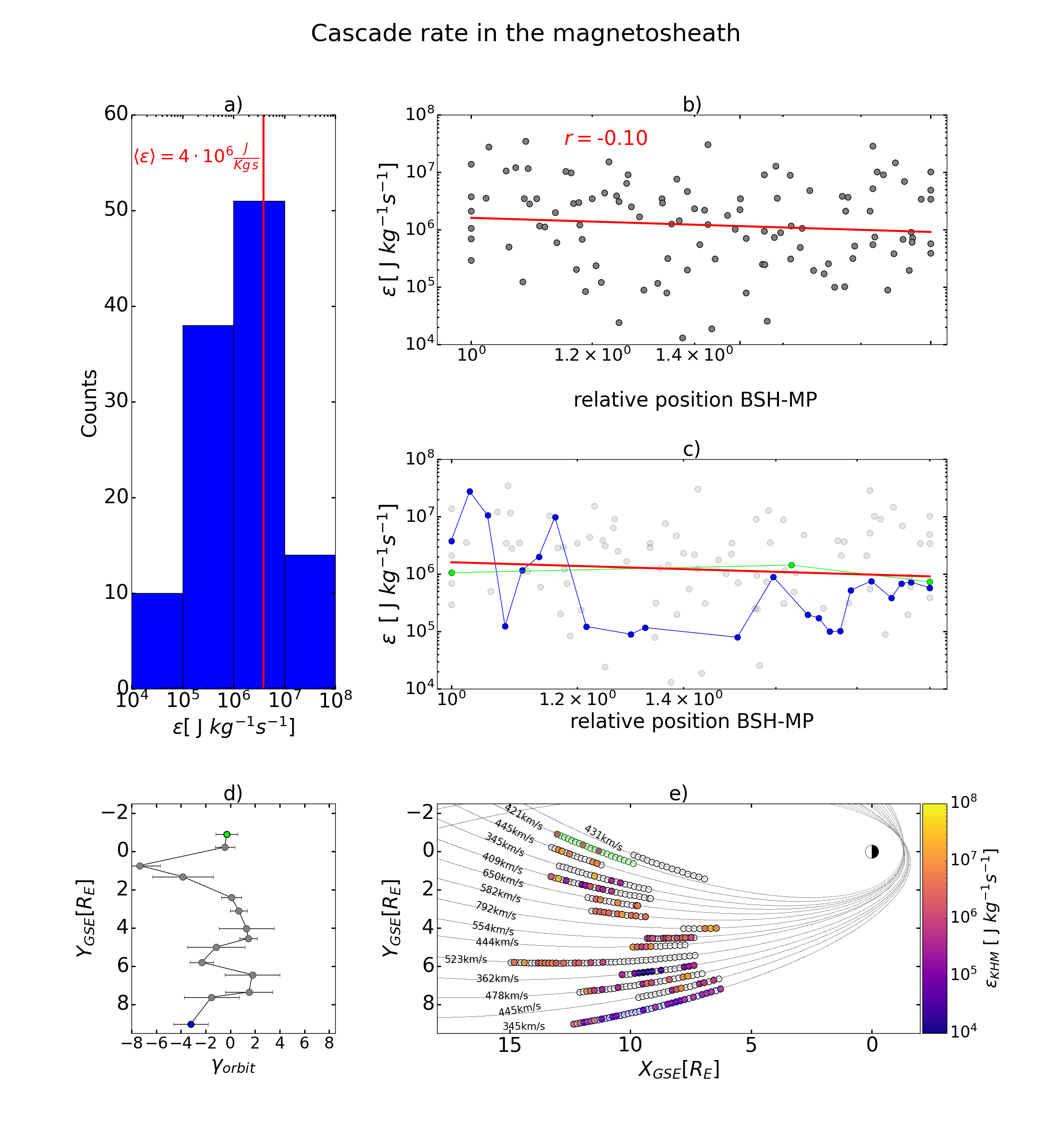}
\caption{Properties of the KHM cascade rate in the magnetosheath during the unbiased campaign. Panel a): Histogram of the cascade rate values computed from convergent burst intervals of the unbiased 
campaign. 
The vertical red line indicates the mean value. 
Panel b): Evolution of the KHM cascade rate, $\varepsilon$, with respect to the relative distance 
to the bow shock $D_{B}$ for all convergent intervals. In order to represent $\epsilon$ versus $D_{B}$ in log-log 
scale, we set the bow shock crossings at $D_{B}=1$ and the magnetopause crossing 
at $D_{B}=2$. Panel c): The same as panel b), but with the data points from the closest magnetosheath traversal to the flanks (orbit 1 in figure\ref{fig_orbit}) highlighted in blue and the closest ones to the subsolar region with convergent burst intervals in green (orbit 14 in figure\ref{fig_orbit}). Red lines in 
panel b) and c) correspond to a power-law fit of $\varepsilon$ and $D_{B}$ computed for all convergent time intervals, and the correlation coefficient is indicated as $r$ in panel b). Panel d): exponent of the power-law fit $\varepsilon= n\cdot D_{B}^{\gamma}$ for the convergent intervals of each magnetosheath traversal versus 
the location in $Y_{GSE}$ of MMS1 at $D_B = 1$. The blue and green points correspond to orbits 1 and 14 of the campaign, 
respectively. Panel e): location of MMS1 in the XY GSE plane at the middle of each burst interval.  
Colored circles indicate the KHM cascade rate of the convergent intervals. Non-convergent intervals are empty. The contour of the circles in orbits 1 and 14  
are highlighted in blue and green. 
  Velocities in the left of each magnetosheath crossing indicate the averaged upstream solar wind speed corresponding to each magnetosheath traversal.
}

\label{fig_KHM_cascade}
\end{figure}

Figure 
\ref{fig_KHM_cascade} contains all the information related to the cascade rate 
measurements in the magnetosheath during 
the unbiased campaign. Panel \ref{fig_KHM_cascade}a) shows the histogram of the cascade rate for the convergent intervals. The measured cascade rates have values that span several orders 
of magnitude, with about
half of them in the range $ 10^{6}-10^{7}$ [J\,kg$^{-1}$s$^{-1}$]. The average value, marked as a vertical red line in panel 
\ref{fig_KHM_cascade}a), is $4\cdot 10^{6}$ [J\,kg$^{-1}$s$^{-1}$]. This average value 
is compatible with previous 
estimates of the energy cascade rate in the 
magnetosheath \citep{2018PhRvL.120e5102H,bandyopadhyay_incompressive_2018,2023PhRvL.131v5201P}.

Panel \ref{fig_KHM_cascade}b) shows how the convergent intervals are placed relative to 
the bow shock and the magnetopause, $D_{B}$. We remark that, in order to visualize 
the data in log-log scale in panels \ref{fig_KHM_cascade}b) and \ref{fig_KHM_cascade}c), we 
placed the bow shock crossing at $D_{B}=1$ and the magnetopause crossing at $D_{B}=2$. We fit the measured cascade rates to a power law function with $D_{B}$ of the 
form $\varepsilon=n D_{B}^{\gamma}$. The correlation coefficient of this fit is 
$r=-0.10$, weaker than the ones found for the linear fits of the spectral indices 
with $D_{B}$.

In panel \ref{fig_KHM_cascade}c) the cascade rates in crossing 14 (green points), the closest to the subsolar region with convergent intervals, and crossing 1 (blue points), the closest to the magnetosheath flanks,   
show a decrease with the distance to the bow shock. A similar behavior can be 
appreciated in other magnetosheath crossings. Panel \ref{fig_KHM_cascade}d) shows 
the position of MMS1 in $Y_{GSE}$ at the moment of the bow shock crossing in relation 
to the exponents of the power law fits, $\varepsilon=n D_{B}^{\gamma}$, for each 
magnetosheath traversal. Seven out of the fifteen magnetosheath traversals of the 
campaign present negative exponents, which indicate a decay of the cascade rate 
as the distance to the bow shock increases, 
similar to how the cascade rate decays with heliospheric distance in the solar wind  
\citep{Sorriso-Valvo2023}. In the other cases, the exponents were 
positive or close to zero. 
As can be seen in table \ref{table_1}, the 
magnetosheath crossings that display this behavior are, in general, those with slow upstream plasma speeds.
In panel \ref{fig_KHM_cascade}e), it can also be observed 
that they correspond to the shortest magnetosheath traversals of the campaign. It 
should be remarked that there are less convergent intervals near the 
subsolar region. This is mostly caused by Taylor's condition $U>v_A$ not being 
valid in many burst intervals of this region of the magnetosheath. On the other 
hand, there is no clear relation between the presence of a convergent interval 
and its location with respect to the bow shock. For four magnetosheath crossings, 
convergent burst intervals can only be found near the bow shock, while in other 
three crossings they are mostly located close to the magnetopause. For the other 
magnetosheath traversals, the convergent intervals are distributed along 
spacecraft orbit with no particular ordering. We have not found a link between 
how the convergent intervals are distributed in a magnetosheath crossing and the 
plasma properties upstream or downstream the bow shock.  
These non-convergent intervals correspond to the empty circles in panel 
\ref{fig_KHM_cascade}e).

\begin{figure}
\includegraphics[width=0.5\textwidth]{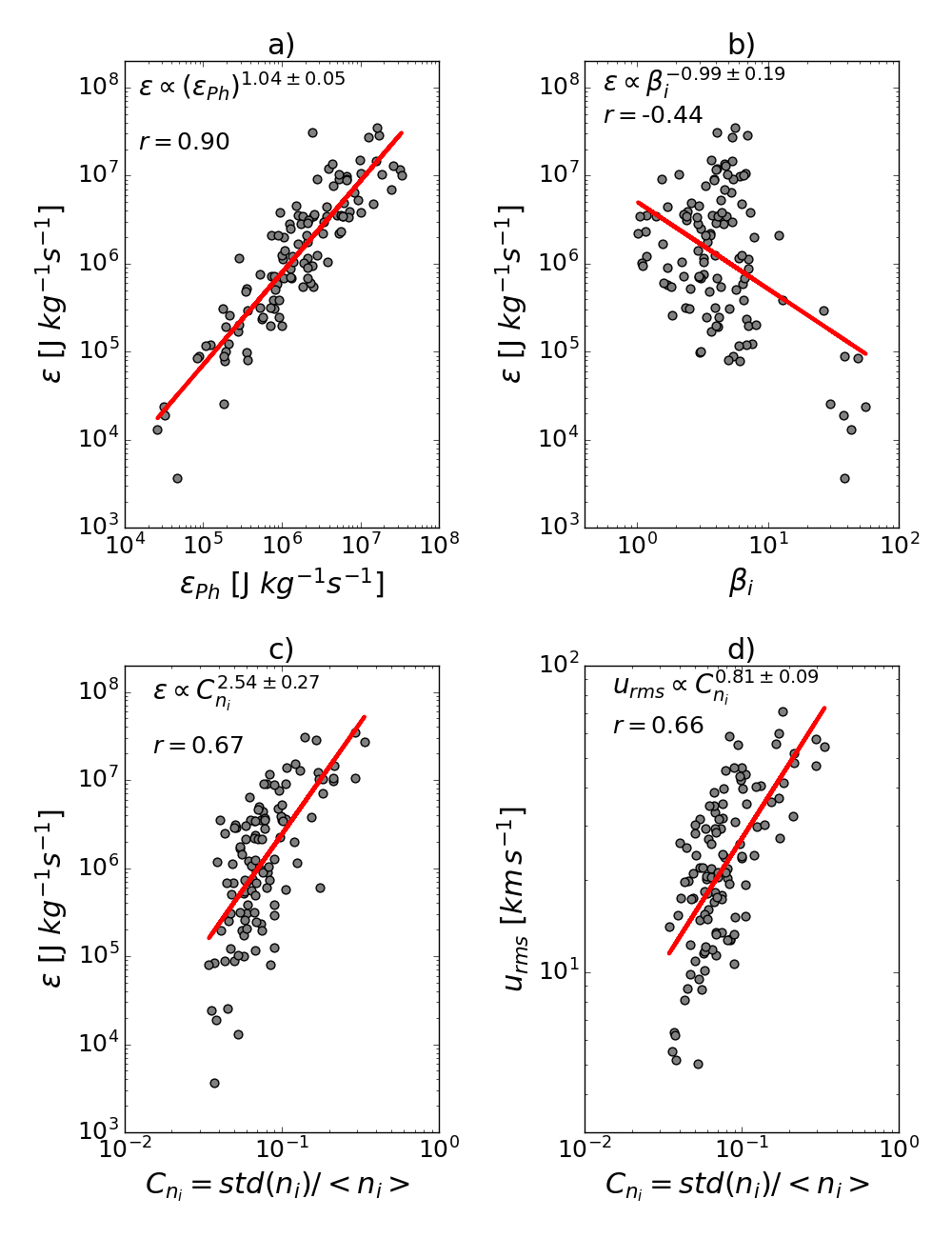}
\caption{ Correlation of the KHM cascade rate with a) 
phenomenological estimate of the cascade rate, $\varepsilon_{Ph}$, as defined in equation \ref{eq_vK}, b) the ion plasma beta $\beta_i$, c)  
the normalized density fluctuations $C_n=std(n_i)/\langle n_i\rangle$ (where std stands for standard deviation), and d) the correlation of $C_n$ 
with the root mean square of velocity fluctuations $u_{rms}$. Grey circles correspond to convergent burst intervals of the campaign.
}
\label{fig_cor_eps}
\end{figure}

The cascade rate estimated with the KHM equation \ref{eq_exact_law}, 
$\varepsilon_{KHM}$, correlates better with other magnitudes different from the 
distance to the bow shock. In figure \ref{fig_cor_eps} we present some plasma 
parameters that correlate well with $\varepsilon_{KHM}$. Panel \ref{fig_cor_eps}a) 
compares $\varepsilon_{KHM}$ with the 
estimates of the cascade rate given by 
a phenomenological formula based on the amplitudes of the 
magnetic and velocity fluctuations, and the correlation lengths \citep{1995PhFl....7.2886H,1999PhRvL..82.3444M,2012JFM...697..296W,2018ApJ...866...81B}

\begin{eqnarray}
\varepsilon_{Ph} = C^{+}\frac{|z^{+}_{rms}|^2z^{-}_{rms}}{4\ell^{+}_{corr}}+C^{-}\frac{|z^{-}_{rms}|^2z^{+}_{rms}}{4\ell^{-}_{corr}}
\label{eq_vK}
\end{eqnarray}

where $\bs{z}^{\pm}=\bs{u}\pm\bs{b}$ are the Elsässer variables, $a_{rms}$ 
denotes the root 
mean squared amplitude of a generic quantity $\bs{a}$, $\lambda^{\pm}$ are the 
correlation lengths of the Elsässer variables and $C^{\pm}$ are the von Karman experimental constants. 
Following the reasoning in \citet{2018ApJ...866...81B}, we assume $C^{+}=C^{-}=0.03$ and that the Elsässer correlation length are equal to the 
correlation length of the magnetic fluctuations, that is, $\ell^{+}_{corr}=\ell^{-}_{corr}=\ell_{corr}$. Both, the KHM and the phenomenological  
estimation of the cascade rate, fit a power law 
$\epsilon_{KHM}\propto\epsilon_{Ph}^{1.04\pm 0.05}$ with correlation coefficient $r=0.90$, which is indicative of $\epsilon_{Ph}$ being a good estimation of $\epsilon_{KHM}$ in the region of the magnetosheath sampled during the 2023 unbiased campaign.

Panel b) in figure \ref{fig_cor_eps} shows that the KHM cascade rate fits a power law 
with the ion plasma beta $\beta_i$, although with weaker correlation than $\epsilon_{KHM}$, $r=-0.44$. A power-law fit of the 
alfvénic Mach number presents the same correlation coefficient as $\beta_i$ (not shown). In 
panel \ref{fig_cor_eps}c) we can observe that the cascade rate has a positive correlation 
with the ion density fluctuations, computed as $C_n=std(n_i)/n_i$, where $std(n_i)$ is the 
standard deviation of the ion number density $n_i$. The power-law fit of $
\varepsilon_{KHM}$ with  $C_n$ has a correlation coefficient of $r=0.67$. 
Panel \ref{fig_cor_eps}d) shows that $u_{rms}$ is proportional to a power law of $C_n$, with $r=0.66$. These results indicate that a highly 
compressible plasma, which is characterized by large density fluctuations, is also 
accompanied by large fluctuations of the plasma speed. This, in turn, implies a higher 
cascade rate due to the dependence of 
$\varepsilon_{Ph}$ on $u_{rms}$. 
The strong correlation of $\varepsilon_{Ph}$ and $\varepsilon_{KHM}$ 
confirms that, this effect of plasma compressibility also has an impact on the KHM estimations of the cascade rate. In summary, our results suggest  
that the cascade rate benefits from the increase of compressibility in the magnetosheath, 
in agreement with previous studies on compressible plasma turbulence \citep{2018PhRvL.120e5102H}.

\subsection{Effect of upstream conditions} 
\label{sec:SW}

When solar wind plasma crosses the bow shock, turbulence fluctuations are altered. We 
evaluate this change in the fluctuations upstream and downstream the bow shock for each 
magnetosheath traversal of the unbiased campaign and determine its effect on the energy 
cascade rate. In order to do so, we use WIND spacecraft measurements to compute the average 
solar wind bulk speed $U_{SW}$, the root mean square of the velocity and the magnetic field fluctuations, 
$u_{rms,SW}$ and $b_{rms,SW}$, 
the correlation length associated to the magnetic fluctuations and the  phenomenological cascade rate $\varepsilon_{Ph,SW}$. 
These averaged quantities are computed over time intervals 
starting one hour prior to each  
MMS bow shock crossing of the campaign. Since the correlation time in the solar wind can typically range from one 
to three hours, and therefore be longer than some of the magnetosheath traversals 
in the unbiased campaign, we chose to compute the correlation length and the average values of the aforementioned solar wind properties in intervals with a duration of 8 hours, rather than just the duration of their corresponding MMS magnetosheath crossing. Solar wind properties remain stable during these 8-hour intervals. 
We measure the same quantities for each convergent burst interval 
in the magnetosheath and take the average of the convergent intervals in   
each magnetosheath crossing. We also evaluate the ratios 
$u_{rms,MSH}/u_{rms,SW}$, $b_{rms,MSH}/b_{rms,SW}$, $\ell_{corr,MSH}/\ell_{corr,SW}$, and $\varepsilon_{Ph,MSH}/\varepsilon_{Ph,SW}$.
Finally, we compute the average KHM cascade rate of all convergent intervals in each magnetosheath 
crossing. Table \ref{table_1} contains all these quantities, as well as the percentage of convergent 
intervals in each magnetosheath traversal.

The results in table \ref{table_1} show that the deceleration of the several hundred 
km/s that experiences the upstream plasma after crossing the bow shock is generally 
accompanied by an increase of  
magnetic fluctuations and the reduction of the correlation length.
On average, velocity 
fluctuations change by a factor of $0.9\pm 0.7$ once the solar wind plasma crosses the bow 
shock and enters the magnetosheath. Magnetic field fluctuations, on the other 
hand, are enhanced by a factor $1.3\pm 0.9$, and the correlation rate in the 
magnetosheath becomes three orders of magnitude smaller than in the solar wind, $(2\pm 2)\,10^{-3}$. 
In turn, the cascade rate increases two orders of magnitude when the plasma crosses the bow shock, $\varepsilon_{Ph,MSH}/\varepsilon_{Ph,SW}\approx 400$. If we take the average cascade rate in the magnetosheath computed with 
the KHM equations $\varepsilon_{KHM,MSH}$, $(4\pm 6)\,10^{6}$ J\,kg$^{-1}$s$^{-1}$ and compare it to the $\varepsilon_{KHM,SW}$ measured in other studies 
\citep{2008ApJ...679.1644M,2009ApJ...697.1119S,coburn2015third}, we find that the 
ratio is about two to three orders of magnitude and is therefore compatible with the 
ratio of the phenomenological estimation of the cascade rate.

\section{Discussion} 
\label{sec:discussion}
\subsection{Spectral index evolution with bow shock}
\label{sec:disc_spectra}

Our results indicate a change of the spectrum of magnetic field fluctuations with respect 
to the relative distance to the bow shock, $D_B$, that suggests an evolution of turbulence 
from an early formation stage near the bow shock towards a well-developed cascade in the 
proximity of the magnetopause. The steepening of the inertial range spectrum that we 
observe, from a power law $f^{-1}$ towards one close to $f^{-1.8}$, and the flattening of the spectrum in the sub-ion range scales, which approach 
$f^{-2.9}$ as $D_B$ increases, follows the same trend than the spectra in the solar wind as heliospheric distance increase \citep{2020ApJS..246...53C,2023ApJ...943L...8S,2020ApJS..246...55D,2022A&A...668A.189P,2025ApJS..278...40S}. This suggests that turbulence in the magnetosheath undergoes a similar evolution than in the solar wind at a faster rate, as it reproduces the same 
behavior in an smaller system. This trend is less apparent for the magnetosheath 
crossings near the subsolar region, where the travel path is shorter and spectra appear less altered by the bow shock crossings, presenting minimal change with $D_B$ and  spectral indices closer to those generally observe in the solar wind at 1 AU.

These changes of the spectral indices at inertial and sub-ion scales and their dependence 
to $D_B$ are in agreement with a statistical study based on Cluster mission observations  
\citep{huang_existence_2017}. There, \citet{huang_existence_2017} determined the 
spectral slope in the inertial and sub-ion range of 1600 time intervals of 1.6-hours   
duration and a sampling rate of 23 Hz. They showed that at inertial range scales, 
spectral indices were flatter near the bow shock, and at sub-ion scales, the spectra were 
less steep close to the magnetopause. The same trends in the evolution of the inertial and sub-ion range spectral indices in the magnetosheath were reported by \citet{2020ApJ...898L..43L} based on an analysis of 1841 selected segments of burst intervals from  MMS1. Our results present the same dependence of the 
spectral indices with the distance to the bow shock and with the distance to the subsolar 
region. The unbiased dataset used in this work allows to confirm that this evolution 
with distance of the spectral indices remain true for individual magnetosheath crossings 
and that it is independent of any selection criteria for the intervals.

\citet{huang_existence_2017} also found that the distribution of inertial range spectral 
indices in the magnetosheath peaks near $-1.2$, instead of the $-1.5$ found in our work. 
This discrepancy could be caused by a selection bias of the 
intervals in \citep{huang_existence_2017}, something that can be discarded from our 
dataset. However, it could also be the result of the limited coverage of the 2023 
unbiased campaign. This study is focused on a region of the magnetosheath where the 
shock normal and the upstream magnetic field form a quasi-perpendicular angle 
\citep{2024JGRA..12932661S} and only a limited number of upstream solar wind conditions 
are available during the 2023 unbiased campaign. In the future, we will extend the 
present analysis to the 2024 and the 
 2026 MMS unbiased campaigns, which will 
provide a broader distribution of shock angles and solar wind conditions. 
The 
data from these campaigns will allow for a spatial coverage comparable 
to previous surveys of turbulence properties in Earth's magnetosphere 
\citep{huang_existence_2017,2020ApJ...898L..43L}, with the advantages 
of an unbiased dataset. The incorporation of measurements from the 
new unbiased campaign observations into our analysis will allow to 
determine the origin of this difference in the spectral index 
histograms at inertial range scales.

\subsection{Enhancement of the cascade rate across the bow shock}
Our measurements of the energy cascade rate via the equation \ref{eq_exact_law} show 
that the cascade rate in the magnetosheath can be several orders of magnitude larger 
than in the solar wind. The average cascade rate computed using the unbiased 
magnetosheath campaign is  
$<\varepsilon_{KHM}>=(4\pm 6)\cdot10^{6}$ J\,kg$^{-1}$s$^{-1}$, larger than similar 
measurements done in the solar wind at 1 AU, which span between 
$10^{3}$ - $10^{4}$ J\,kg$^{-1}$s$^{-1}$ \citep{2008ApJ...679.1644M,2009ApJ...697.1119S,coburn2015third}. This result is in line with previous studies that have 
quantified the cascade rate in Earth's magnetosheath with the KHM equations under the assumptions of plasma incompressibility and turbulence isotropy 
\citep{2018PhRvL.120e5102H,bandyopadhyay_incompressive_2018,2020PhRvL.124v5101B} and incompressible KHM  with no assumptions on anisotropy 
\citep{2023PhRvL.131v5201P,2026A&A...708A.208P}.

Attending to the strong correlation that $\varepsilon_{KHM}$ has with the 
phenomenological estimation of the cascade rate, we used the 
latter to measure the energy cascade rate upstream of the bow shock 
during each magnetosheath traversal of the unbiased campaign. We observed that the 
average increase of the ratio 
$\varepsilon_{Ph,MSH}/\varepsilon_{Ph,SW}$ is about 400, in line with the 2 - 3 orders of magnitude obtained from the KHM estimation of the cascade rate. 
 This increase in the cascade rate downstream of the 
bow shock corresponds only to measurements taken in Earth's 
magnetosheath. Recent works have shown that the cascade rate can present important variations in different regions of the near-Earth environment. \citet{2026PhPl...33a2903Z} obtained that 
the cascade rate in Earth's magnetotail is an order of magnitude larger than in the magnetosheath. \citet{2023ApJ...945....8A} on the other hand, found that the cascade rate in the Jovian magnetosheath is two orders of magnitude smaller than in Earth's one, thus showing that turbulence transfer rate can also be substantially different in other planetary magnetosheaths. This variety of   cascade rate values is easier to interpret in terms of changes in plasma properties by using a phenomenological description of the energy transfer rate. Assuming low correlation of the velocity and magnetic fluctuations, the phenomenological cascade rate estimation, $\epsilon_{Ph}$, given in equation \ref{eq_vK} can be approximated as 

\begin{eqnarray}
\varepsilon_{Ph} \approx C^{+}\frac{(u_{rms}^2+b_{rms}^2)^{3/2}}{2\ell_{corr}}
\label{eq_vK2}
\end{eqnarray}

This simplified expression evidences the impact of the velocity and 
magnetic fluctuations. As in other interplanetary shocks, the crossing of the bow 
shock produces an increase in the level of the fluctuations \citep{2017ApJ...844...51P}, which contributes to the enhancement of the cascade rate. This 
increase of the energy transfer rate caused by the large amplitude fluctuations 
in the magnetosheath is complemented with the decrease of the correlation length 
in the same region, which has a substantial effect on the energy cascade.

Table 1 shows that if the upstream plasma is a fast wind, the change of the 
correlation length after crossing the bow shock, given by the ratio $\ell_{corr,MSH}/\ell_{corr,SW}$, is smaller than for slow winds. This 
dependence of the correlation length with the upstream conditions has already 
been observed in Earth's magnetosheath \citep{2009JGRA..114.8207G}. 
\citet{2026JGRA..13134919P} also reported a small growth of the correlation 
length with increasing distance from  the bow shock during the 2023 unbiased 
campaign, which is in accordance with other statistical surveys of the 
correlation length in  Earth's magnetosheath \citep{2008AnGeo..26.2503G,huang_existence_2017,2022PhPl...29a2302S}. This behavior of the 
magnetosheath's correlation length with the distance to the bow shock translates 
only slightly to the cascade rate, which displays a small decay with distance  for the slowest winds, specially near the flank region.

As shown by equation \ref{eq_vK2}, the effect of the amplitude of the 
fluctuations on the cascade rate can alter the value of $\varepsilon$ across the 
magnetosheath. It is left to determine how the cascade rate will evolve in a 
different region of the magnetosheath where the level of the fluctuations can have an even larger impact on the transfer of energy, for instance, in a region downstream a quasi-parallel shock. Since fluctuations downstream quasi-parallel shocks have generally 
larger amplitude than the quasi-perpendicular ones \citep{1986JGR....91.1711L,2025JGRA..13033272S}, we expect that under the 
same upstream conditions, the cascade rate in quasi-parallel 
magnetosheath will also be larger than in the quasi-perpendicular region. However, this assumption relies on turbulence in the quasi-parallel magnetosheath also having a good agreement with 
the phenomenological estimation of the cascade rate  
defined in equation \ref{eq_vK}. It is possible that, unlike for the 2023 unbiased campaign, the assumptions made on equation \ref{eq_vK} about the correlation lengths, $\ell^{\pm}_{corr}$, and the Kolmogorov constants, $C^{\pm}$, lead  to a estimates of the cascade rate in the quasi-parallel magnetosheath that do not correlates as well with the KHM measurement of the energy transfer, $\varepsilon_{KHM}$. The changes on the cascade rate estimations in the magnetosheath downstream the quasi-parallel bow shock will be included in a future study.

\subsection{Cascade rate evolution across the magnetosheath}
The computation of the KHM energy cascade rate, $\varepsilon_{KHM}$ has allowed to 
evaluate the dependence of $\varepsilon_{KHM}$ with the relative distance to the bow 
shock, $D_{B}$. 
The $\varepsilon_{KHM}$ measured for all convergent intervals do 
not correlate well with $D_{B}$. However, when looking at the evolution of the cascade rate in each magnetosheath traversals, we observe that 
$\varepsilon_{KHM}$ decays as a power law of $D_{B}$ in 7 of the 15 inbound magnetosheath 
crossings. This power-law decay of the cascade rate with distance is qualitatively 
similar to that observed in the solar wind \citep{Sorriso-Valvo2023}. For the other 8 magnetosheath crossings, the power-law exponent with $D_{B}$ is 
positive or close to zero in 7 of them. In the remaining one, corresponding to crossing 15 in table \ref{table_1}, the KHM cascade rate could not be computed since none of the intervals fulfilled Taylor's hypothesis. The upstream wind conditions seem to play a role in the 
evolution of $\varepsilon_{KHM}$ with $D_{B}$, as the magnetosheath traversals where the cascade rate decays are generally the ones with lower upstream wind speed.  
In the future, with more data available from other magnetosheath campaigns and a larger sample of upstream conditions, we will further study the influence of the upstream wind speed on the evolution of the energy cascade rate across the magnetosheath.

The possible link between upstream conditions and $\varepsilon_{KHM}$   
does not explain how, 
for some magnetosheath crossings, the cascade rate increases as the plasma 
moves away from the bow shock. As mentioned in the previous section, the bow shock 
injects large amplitude fluctuations to the solar wind plasma as it enters the 
magnetopause, thus acting as a source of energy for turbulence. One would expect  
that, as the distance from the bow shock increases, the cascade rate would decay.

Another unexpected result is that, close to the bow shock, we found intervals with a 
constant non-linear transfer of energy across scales but with spectra that did not 
present a Kolmogorov's $f^{-5/3}$ power law. In figure \ref{fig_KHM_example}, we show an 
example of a magnetosheath interval that displays a constant non-linear energy transfer 
and a spectrum that could be associated to an underdeveloped turbulence. 
As explained  in section \ref{subsec:psd}, the presence of coherent structures or wave activity can produce local enhancements in the spectrum of fluctuations, particularly near the transition region. These enhancements can lead to the flattening of the spectrum above sub-ion scales or to the formation of bumps. There are, however, other mechanisms that could explain the formation of a turbulence cascade in an interval with a seemingly underdeveloped turbulence spectrum, or could even clarify the rise of $\varepsilon_{KHM}$ with increasing distance from the bow shock that we observe in half of the magnetosheath crossings. These mechanisms are related to the several assumptions we made to provide an estimate of the cascade rate.

Equation \ref{eq_exact_law}, which we used to measure the 
KHM cascade rate, requires the assumption
 that the plasma in the magnetosheath can be described with the incompressible Hall-MHD equations and that turbulence is isotropic. 
Taking into account turbulence anisotropy or compressible effects has proven to 
substantially increase the estimated cascade rate \citep{2009ApJ...697.1119S,2015ApJ...804..119V,2018PhRvL.120e5102H,2019PhRvL.123x5101A} and works like  \citet{2021JGRA..12628780H} and \citet{2024A&A...684A.120H} pointed out the need to take into account turbulence anisotropy even at sub-ion scales. Recent studies also indicate that 
other effects, such as plasma expansion, might be an important contribution to the 
energy transfer, specially for intervals with a $f^{-1}$ spectrum \citep{2024A&A...690A.265V}. If all these effects dominate the cross-scale 
transfer of energy near the bow shock, a cascade rate estimation that accounts 
for all of them  
would present a more clear decay with distance, even for the crossings 
where this behavior is not observed when incompressibile isotropic turbulence is 
assumed. The intervals with flatter spectra would also be the ones mostly affected by  
contributions such as compressibility, as they are measured generally close to 
the bow shock. Hence, the study of these contributions could help reveal the nature of 
the $f^{-1}$ spectrum in the magnetosheath. We aim to evaluate the impact of 
compressibility and anisotropy on the cascade rate and their dependence on the 
magnetosheath location in a future study.

\section{Conclusions}
\label{sec:conclusions}
We have computed the energy cascade rate and analyzed the spectra magnetic field 
fluctuations of 309 high cadence data intervals gathered by the MMS spacecraft during the 2023 MMS unbiased magnetosheath campaign. The time intervals, taken during 15 consecutive 
inbound magnetosheath crossings, were evenly distributed from the bow shock to the 
magnetopause, each of them with a duration of 3 minutes and separated by a 6 
minute gap from each other. No other 
criteria were imposed to select these high cadence intervals. These properties of the unbiased campaign dataset 
have allowed us to measure the cascade rate and the power spectra of magnetic field  
fluctuations in the magnetosheath without a selection bias and to take these measurements from the inertial range to scales well into the sub-ion range. 
We summarize our findings as follows:

\begin{itemize}
\item The change of the spectral 
indices in the magnetosheath at inertial and sub-ion scales with the distance to the bow shock is similar to that observed in the solar wind, and is therefore compatible with an evolution of turbulence towards well-developed state as it approaches the magnetopause. 
The spectral index in the inertial range is positively correlated with the relative distance to the bow shock. The slope steepens from a power law close to $f^{-1}$ near the bow shock to one closer to $f^{-1.8}$. At sub-ion scales, the spectral index presents a weaker negative correlation with distance. It goes from a steep spectrum of 
$f^{-3}$ near the bow shock, towards spectral indices closer to 
$-2.9$ near to the magnetopause. We observe that close to the sub-solar region, the spectral indices at the bow shock are closer to the ones of a well-developed turbulence and their evolution with distance is minimal.

\item We computed the cascade rate in the unbiased campaign intervals using a KHM 
equation derived from incompressible Hall-MHD. 
After visual inspection, about half of 
the initial burst intervals presented a constant energy transfer across scales. 
In addition, we discarded the 
intervals that did not fulfill Taylor's hypothesis, leaving 114 of the 
initial 309 burst intervals where the cascade rate has been computed.
The average cascade rate during the campaign, 
$(4\pm 6)\,10^{6}$ J\,kg$^{-1}$s$^{-1}$ is two to three orders of magnitude 
larger than typical values of the KHM cascade rate measured in the solar 
wind at 1 AU. We found that the cascade rate computed with the KHM equation has a 
strong correlation with a phenomenological estimation of the cascade rate computed with the amplitudes of the velocity and magnetic fluctuations, and the correlation length of magnetic fluctuations. Using measurements of these fluctuations and correlation lengths upstream and downstream the bow shock, we obtained the phenomenological estimate of the cascade rate 
in both regions. We found that, on average, the cascade rate increases by a factor close to 400 once the 
plasma crosses the bow shock into the magnetosheath. Taking into account the quantities 
involved in the phenomenological formula, we attribute the enhancement of the cascade rate 
to the increase in amplitude of the fluctuations and the decrease of the 
correlation length that takes place after the solar wind plasma crosses the bow 
shock and enters the downstream region.

\item Unlike the spectral indices, the cascade rate computed from the KHM equation
only presents a weak correlation with the distance to the bow shock. 
However, for half of the magnetosheath crossings, with upstream speed generally below 450 km/s, 
the KHM cascade rate decayed following a power law with the distance to the bow 
shock. For the other magnetosheath crossings, the cascade rate remained almost 
constant or increased with distance. Since the injection of energy is produced at the bow 
shock crossing due to the enhancement of the fluctuations arriving from the upstream 
region, the increase of the cascade rate as the plasma moves away from the source of 
energy is an unexpected result. This behavior of the cascade rate is also 
at odds with the decrease of the correlation length with distance to the bow 
shock reported in previous studies 
\citep{2009JGRA..114.8207G,2026JGRA..13134919P}. We conjecture that contributions to the cascade rate not 
considered in this work, such as compressibility, expansion or turbulence anisotropy, 
might play a major role in these magnetosheath crossings and, once implemented into the 
estimation of the 
cascade rate, it will also present a decay with increasing 
distance from the bow shock. 
\end{itemize}

The 2023 unbiased campaign observations have allowed us to analyze 
turbulence properties across multiple magnetosheath crossings, from large to sub-ion scales, 
in a dataset that is not influenced by any selection bias. Yet, the launch of 
new MMS unbiased campaigns is still 
necessary to have a completely unbiased view of the magnetosheath with 
high-resolution measurements. In this work we have remarked that plasma conditions 
upstream the bow shock can affect certain turbulence properties in the magnetosheath 
and their evolution as the plasma approaches the magnetopause. Increasing the number 
of unbiased magnetosheath crossings with high-cadence sampling would allow to 
find what solar wind properties have a greater influence on the magnetosheath 
and could be used to determine whether our results can be extended to other magnetosheath 
regions under different upstream conditions. Ultimately, a deeper understanding of 
Earth's magnetosheath turbulence will also contribute to gain a better insight into the dynamics of 
turbulence downstream interplanetary shocks and how they change under different ambient plasma 
conditions.

\begin{acknowledgments}
We acknowledge support of Grant PID2023-147331OB-I00 funded by 
MICIU/AEI/ 10.13039/501100011033 and FEDER, UE. Part of this publication is 
also supported by the I+D+i project PID2023-147708NB-I00 funded by
MICIU/AEI/10.13039/501100011033/ and  by  FEDER, EU. 

V.M.-C. is supported by the Vicenç Mut grant POSTDOC2024\_58 funded by 
Conselleria d'Educació i Universitats del Govern de les Illes Balears i pel 
Fons Social Europeu +. P.H. acknowledges grant 25-17802S of the Czech Science Foundation. L.S.-V. was funded by the Swedish National Space Agency (SNSA) Research Grant 2025-00229 and by the PRIN274 2022 project “2022KL38BK - The ULtimate fate of TuRbulence from space to laboratory plAsmas (ULTRA)” (Master CUP B53D23004850006) by the Italian Ministry of University and Research, funded under the National Recovery and Resilience Plan (NRRP), Mission 4 – Component C2 – Investment 1.1, “Fondo per il Programma Nazionale di Ricerca e Progetti di Rilevante Interesse Nazionale (PRIN 2022)” (PE9) by the European Union – NextGenerationEU. F.P. is supported by NSF SHINE 2501387, NASA MMS Mission Theory, Modeling and Data Analysis Program under grant number 80NSSC19K0565 at the University of Delaware, and a Plan for NASA EPSCoR Research Infrastructure Development (RID) in Delaware (NASA award 80NSSC22M0039). JES is supported by the Royal Society University Research Fellowship URF\textbackslash R1\textbackslash 201286 and URF\textbackslash R\textbackslash 251029.

All data used in this work is publicly available. MMS data can be found at the MMS Science Data Center (https://lasp.colorado.edu/mms/sdc/public). We thank the MMS team and instrument leads for the data access and support. We acknowledge the use of NASA/GSFC's Space Physics Data Facility's OMNIWeb to retreive data from WIND mission.
\end{acknowledgments}

%



\software{We acknowledge the use of the pyrfu analysis package \citep{richard_2024_10678695} 
 and the pyspedas library \citep{angelopoulos_2026_21812086}.}

\bibliography{unbiased.bib}
\bibliographystyle{aasjournalv7.1}

\begin{longrotatetable}
\begin{deluxetable*}{ccccccccccc}
\tablecaption{
Average values for each magnetosheath crossing during the 2023 MMS unbiased 
magnetosheath campaign. The last row, shows the mean values and standard deviations 
computed from all 309 burst intervals. The first column of the table indicates the 
number assigned to each inbound crossing. The second and third 
columns show the start and end times of each crossing. The fourth column is the 
solar wind bulk speed measured by WIND spacecraft and averaged over a time interval with a duration of 8 hours, starting one hour before each magnetosheath crossing.
This 1-hour time shift is done in order to account 
for wind propagation to the bow shock. The fifth column is the average 
plasma bulk speed in the magnetosheath during the MMS traversal, computed from the 
average of all burst intervals in each magnetosheath traversal. Column 6 corresponds to the ratio of the root mean square of the velocity fluctuations in the magnetosheath 
(MSH) and the solar wind (SW). Column 7 shows the same ratio for the root mean square of the magnetic field fluctuations. Column 8 indicates the ratio of 
 the correlation length of magnetic fluctuations in the magnetosheath and in the solar wind. Column 9 corresponds to the average energy cascade rates in the 
magnetosheath and the solar wind computed with 
the phenomenological formula in equation \ref{eq_vK}. Column 10 is the average energy cascade rate of the magnetosheath crossing computed with the KHM equation. Column 11 is the percentage of convergent intervals in a magnetosheath traversal.
The magnetosheath quantities in columns 6 to 10 are computed using only 
\textit{convergent intervals}, that is, the ones that show an approximately constant cross-scale energy transfer and fulfill Taylor's hypothesis.
\label{table_1}
}
\tablewidth{700pt}
\tabletypesize{\scriptsize}
\tablehead{
\colhead{MSH crossing} & \colhead{Start} & \colhead{End} & \colhead{$U_{SW}$ [km/s]} &
\colhead{$U_{MSH}$ [km/s]} & \colhead{$u_{MSH}/u_{SW}$}& \colhead{$b_{MSH}/b_{SW}$} & \colhead{$\ell_{corr,MSH}/\ell_{corr,SW}$} & \colhead{$\varepsilon_{Ph,MSH}/\varepsilon_{Ph,SW}$} &
\colhead{$\varepsilon_{KHM}$ [J(kg\,s)$^{-1}$]} & \% convergents \\
} 
\decimalcolnumbers
\startdata
1 & 2023-02-02T21:02:33 & 2023-02-03T02:38:33 & 344  & 158 &	1.2 & 2.2 & $3\,10^{-3}$ & $2.3\,10^{3}$ & $2.7\,10^{6}$ & 58\\
2 & 2023-02-06T12:24:33 & 2023-02-06T15:09:53 & 445  & 212 & 0.7 & 1.1 &  $4\,10^{-3}$ &$1.8\,10^{2}$ & $2.2\,10^{6}$ & 32\\
3 & 2023-02-09T22:46:53 & 2023-02-10T03:10:53 & 478 & 174 & 1.7 & 1.6 &  $3\,10^{-3}$ &$7.9\,10^{2}$  & $4.4\,10^{6}$ & 37\\
4 & 2023-02-13T12:55:23 & 2023-02-13T15:20:23 & 362 & 172 & 1.3 & 2.4 &  $9\,10^{-4}$ &$2.4\,10^{3}$  & $1.1\,10^{5}$ & 59\\
5 & 2023-02-16T20:52:53 & 2023-02-17T03:40:33 & 523 & 225 & 0.8 & 0.9 &  $3\,10^{-3}$ &$7.2\,10^{1}$ & $5.1\,10^{6}$ & 28\\
6 & 2023-02-20T14:32:33 & 2023-02-20T16:11:53 & 444 & 193 & 0.7 & 1.0 &  $2\,10^{-3}$ &$2.1\,10^{2}$ & $4.1\,10^{6}$ & 42\\
7 & 2023-02-24T03:17:03 & 2023-02-24T04:42:13 & 554 & 267 &	0.8 & 1.6 &  $2\,10^{-3}$ &$3.0\,10^{2}$ & $1.4\,10^{6}$ & 56\\
8 & 2023-02-27T16:33:23 & 2023-02-27T17:29:23 & 792 & 380 &	1.4 & 2.0 &  $5\,10^{-4}$ &$3.1\,10^{3}$ & $1.5\,10^{7}$ & 43\\
9 & 2023-03-03T01:57:03 & 2023-03-03T03:46:53 & 582 & 219  &	0.9 & 1.7 &  $5\,10^{-3}$ &$4.0\,10^{2}$ & $4.6\,10^{6}$ & 69\\
10 & 2023-03-06T14:08:13 & 2023-03-06T15:51:33 & 650 & 269 & 1.0 & 2.0 &  $3\,10^{-3}$ &$8.1\,10^{2}$ & $5.3\,10^{6}$ & 38\\
11 & 2023-03-10T00:58:53 & 2023-03-10T04:30:33 & 409 & 149 & 1.1 & 1.5 &  $2\,10^{-3}$ &$8.6\,10^{2}$ & $3.7\,10^{6}$ & 36\\
12 & 2023-03-13T13:23:33 & 2023-03-13T16:35:03 & 345 & 129  & 1.3 & 1.3 &  $7\,10^{-4}$ &$1.4\,10^{3}$ & $8\,10^{6}$ & 14\\
13 & 2023-03-17T01:04:43 & 2023-03-17T03:04:13 & 445 & 160  & 1.4 & 2.3 &  $4\,10^{-3}$ &$1.5\,10^{3}$ & $7.4\,10^{6}$ & 36\\
14 & 2023-03-20T13:52:23 & 2023-03-20T16:45:43 & 421 & 145  & 0.5 & 0.7 &  $8\,10^{-4}$ &$1.9\,10^{2}$ & $1.4\,10^{6}$ & 15\\
15 & 2023-03-24T04:42:53 & 2023-03-24T06:51:23 & 432 & 103  & 1.0 & 1.7 &  \nodata & \nodata & \nodata & 0\\
\textbf{Total} & \nodata & \nodata & $\mathbf{482\pm 114}$ & $\mathbf{187\pm 59}$ & $
\mathbf{0.9\pm 0.7}$ & $\mathbf{1.3\pm 0.9}$ & $\mathbf{(2.\pm 2)\,10^{-3}}$ 
&$\mathbf{(4.28\pm 0.25)\,10^{2}}$ 
& $\mathbf{(4\pm 6)\,10^{6}}$ & $\mathbf{37}$
\enddata
\end{deluxetable*}
\end{longrotatetable}



\end{document}